\documentstyle[12pt,aaspp4]{article}
 
\lefthead{Hutchings et al.}
\righthead{Spectra of 3C galaxy nuclei}
 
\begin{document}
\title{Spatially resolved spectra of 3C galaxy nuclei}
 
\author{J. B. Hutchings\altaffilmark{1}} 
\affil{Dominion Astrophysical Observatory\\Herzberg Institute of Astrophysics,
National Research Council of 
Canada\\ 5071 W. Saanich Rd., Victoria, B.C. V8X 4M6, Canada} 
\author{S. A. Baum\altaffilmark{1}} 
\affil{Space Telescope Science Institute, 1700 San Martin drive,\\
Baltimore, MD 21218} 
\author{D. Weistrop, C. Nelson\altaffilmark{1}} 
\affil{Department of Physics, University of Nevada,\\
Las Vegas, NV 89154-4002}
\author{M. E. Kaiser\altaffilmark{1}}
\affil{Department of Physics and Astronomy, Johns Hopkins University\\
Balrimore, MD 21218}
\author{R. F. Gelderman\altaffilmark{1}}
\affil{Department of Physics and Astronomy, Western Kentucky University\\
Bowling Green, KY 42101-3576}
 
\authoremail{john.hutchings@hia.nrc.ca, sbaum@stsci.edu,weistrop@nevada.edu,
cnelson@hill.physics.unlv.edukaiser@hunin.gsfc.nasa.gov,gelderman@wku.ed}
\altaffiltext{1}{Observer with the NASA/ESA Hubble Space Telescope through 
the Space Telescope Science Institute, which is operated by AURA, Inc., under
contract NAS5-26555}

\begin{abstract}

We present and discuss visible-wavelength long-slit spectra of four
low redshift 3C galaxies obtained with the STIS instrument on the Hubble 
Space Telescope. The slit was aligned with near-nuclear jet-like
structure seen in HST images of the galaxies, to give unprecedented spatial
resolution of the galaxy inner regions. In 3C 135 and 3C 171,
the spectra reveal clumpy emission line structures that indicate outward 
motions of a few hundred km s$^{-1}$ within a centrally illuminated and 
ionised biconical region. There may also be some low-ionisation high-velocity
material associated with 3C 135. In 3C 264 and 3C 78, the jets have blue
featureless spectra consistent with their proposed synchrotron origin. 
There is weak associated line emission in the innermost part of the jets 
with mild outflow velocity. These jets are bright and highly collimated only
within a circumnuclear region of lower galaxy luminosity, which is not dusty. 
We discuss the origins of these central regions and their connection with
relativistic jets.

\end{abstract}

\keywords{galaxies: active, galaxies: jets}

\section{Introduction and observations}

     The HST snapshot program to obtain broad-band images of 3C radio sources 
revealed much of the small-scale optical structure of these extraordinary objects
(De Koff et al 1996). We noticed particularly that a number of them have small
bright structures which are approximately radially oriented from the nuclei.
These jet-like phenomena lie within an arcsec of the nucleus, are 
usually obvious only on one side, and are approximately aligned with the large-scale
radio structure. The original snapshots are only in one colour (F702W filter)
so little more information was initially available. Since then, many of the 
objects have been observed (by SAB and collaborators) with the WFPC2 ramp 
filters to isolate the emission lines of [O III] or H$\alpha$.

    There are several possible origins of these `jets': they may be true synchrotron
jets like 3C 273 or M87; extended line emission regions activated by the AGN like 
NGC 4151; or line emission from gas flows or star-formation triggered by radio
jets. In order to distinguish these possibilities, and learn more about the nuclei
of radio galaxies, we are carrying out a program of long-slit observations
with the Space Telescope Imaging Spectrograph of the Hubble Space Telescope.

    The observations consist of low dispersion optical spectra taken with a
fairly wide (0.5 arcsec) slit placed along the direction of the jets as seen in the
HST snapshots. The slit length is the full $\sim$50 arcsec field of the CCD,
which greatly exceeds the size of the central regions of interest.
The width of the slit means that the emission line resolution is
determined by the spectrograph and the size of features in the galaxies. It also
means that the spectra are essentially slitless for the inner emission-line region, 
and thus similar to the NGC 4151 slitless spectra of Hutchings et al (1998).

    Table 1 gives the journal of observations and basic information on the first 
four program objects. Exposure details were different for each, depending
on the redshift and the flux levels measured from the snapshot images. The
slit angle was positioned to within 5$^o$ of the `jet' and an acquisition image
used to confirm the orientation.  

    The STIS pixel size in the visible wavelength region is 0.051 arcsec, and the spectral
resolution is $\sim$7\AA~ for G450L and $\sim$14\AA~ for G750L for a point
source. Each pixel samples 2.7 and 4.9\AA ~respectively, in dispersion. All
spectra were taken with several readouts, to eliminate cosmic rays. `Superdark'
frames for the days of observation were derived and used to remove the hot pixels 
that are a significant noise source in these faint targets. Wavelength calibration
exposures were taken with each observation and applied using CALSTIS routines in
STSDAS and the STIS team IDL software. Current flux calibrations were applied.
Using the orientation from the STIS data, the WFPC2 snapshots
were rotated and resampled to match the STIS images, and used as undispersed
templates to measure differential velocities of emission line regions within the 
jets. This is described more fully by Hutchings et al (1998). 

    The sections below describe each of the sources individually, as they
are all significantly different. The measurements are described in more
detail in the first object only, to avoid repetition.
 
\section{3C 135}

 This z=0.12 radio galaxy is a giant elliptical in a spiral-rich cluster. 
There is a bright `H$\alpha$ emission'
region 4" SW of the nucleus along the galaxy minor axis (McCarthy et al 1995). 
The HST snapshot image shows this feature clearly, which resembles a superposed 
small elliptical galaxy. The WFPC2 narrow-band image of [O III], obtained through 
the ramp filter, shows weak, if any, emission from this object, suggesting it 
may have a somewhat different redshift, or low ionisation and thus no [O III]. 
The image does show strong [O III] emission around the AGN, almost aligned 
with the SW object but on the other (NE) side of the nucleus.
Figure~\ref{fig1} shows the WFPC2 and STIS images.
 
The radio structure is FR II (edge-brightened lobes) and large ($>$100") but has 
no radio axis given in many tables in references. However, the map of Leahy et al 
(1997) shows the overall radio 
axis has position angle $\sim75^o$, and there is resolved core structure to 
the W lobe, with a knot 0.3 arcsec from the nucleus. The radio axis is $\sim25^o$ 
from the galaxy minor axis, and thus also from the resolved SW bright feature. 
The inner part of the line emission is more closely aligned with the radio 
nuclear structure.

     The WFPC2 snapshot shows extended flux on both sides of the nucleus.
This is almost all from the H$\alpha$ and [N II] lines, as shown by 
continuum-subtracted images derived from the broad-band snapshots. 
Similarly our spectrum shows no significant  
continuum distributed like the extended line emission features.

    The HST narrowband [OIII] image was compared with the spectral images from 
STIS to determine the velocities of the features. The STIS dispersion direction 
is very close to SE-NW, along the host major axis. The structure is fairly complex 
and is seen to the same extent in the narrow band and the spectral images. 
The overall morphology is consistent with bicones seen roughly sideways-on.

 The slit width of 0.5 arcsec covers the full extent of the emission line 
structure across the slit, so the data are effectively slitless over 
the central region. Line emission is seen of H$\beta$, 
[O III] 4959 and 5007\AA, [O I] 6300\AA, [N II] and
H$\alpha$ blend, and the [S II] lines at ~6720\AA. Figure~\ref{fig2} shows contours
of the emission line features and identifies the main ones for reference.
Figure~\ref{fig3}
shows cross-sections along the slit which illustrate the different emission
line regions and their relative fluxes. Figure~\ref{fig4} shows spectra of
different sections of the spectral image plane.

   The Balmer lines are weak, and have most of their off-nuclear flux 
just SW of the nucleus.
Generally the lower ionisation lines are stronger at this location than elsewhere, 
while the 
region NE of the nucleus is seen more strongly in higher ionisation lines. 
Line fluxes are given in Table 2. There is a region just NE of the nucleus that is
devoid of all lines, also seen in the undispersed image. The continuum cross-section 
in regions away from the emission lines does not show this gap, and it is also
not seen in the continuum image. (Since the off-nuclear signal differs from the 
nucleus (Figure~\ref{fig4}), it contains more than the scattered 
nuclear spectrum.)  The gap may be a region where our line of sight to the
emission-line region is obscured, but this is unlikely in view of the symmetric
continuum cross-section. It is more likely to be a real gap in the emission-line clouds.

 It is possible to measure line feature positions to about 0.2 pixels, which 
is about 45 km s$^{-1}$ in the rest frame of 3C135.
Shifts in the location of individual features in the dispersed and undispersed
images indicate the velocities of different
parts of the ELR. The shifts were measured by two different methods. a) A program
was devised to measure flux peaks along each row of the dispersed and undispersed
images. The relative shifts indicate the velocities of different clouds in the 
emission-line region. While objective and sensitive, this suffers from the inability
to disentangle clouds that overlap either spatially or in velocity, particularly in 
the case of different lines whose relative flux depends on ionisation. b) We matched
the positions between dispersed and undispersed features identified by morphology.
While subjective, this enables more distinction between overlaps and ionisation
differences as well as integration over irregular clumps.

   There was good agreement between the two methods, and the results are shown
in Table 2 and Figure~\ref{vels}. The velocity structure is reasonably complex 
and of course suffers from the finite spatial resolution, even with HST. Velocities
range between +500 and -200 km s$^{-1}$ with respect to the nucleus, on both sides 
of it. In this object, our measurements do not indicate a simple picture of
approach on one side, 
and recession on the other, that we see in NGC 4151 (Hutchings et al 1998). 

The SW side is the one with the resolved radio knot, the H$\alpha$
emission, the faint radio jet, and the closer radio lobe. Generally the
stronger emission line features are approaching on this (SW) side and receding
on the NE side (where the higher ionisation lines are stronger).

These results are consistent
with viewing of a bicone nearly perpendicular to its axis, 
although the preponderance of positive cloud velocities with respect to 
the nucleus must be explained.  One possibility is a systematic error in the
nuclear velocity that is the zero point of the distribution. Independent measures
agree well, but a systematic error may result from the crowded nuclear region 
in the data. However, there may be a real difference in the activity on  
opposite sides of the nucleus.
Finally, some models of AGN suggest that the emission-line clouds contain 
significant amounts of dust, and that most of the light we observe is reflected
by the illuminated side of the clouds.  At the viewing angle for this 
object, we would observe more clouds from the `far' side of the cones, which is
receding from us.

Potentially there is also a weak feature with very high apparent
recessional velocity seen in the Halpha + [NII] blend and also in [S II]
and H$\beta$, but not [O III]. 
The feature can be seen in Figure~\ref{fig1}, and appears as
a diagonal rising to the right from below the nucleus to just above it.
The strongest such signal is in the H$\alpha$ blend, but it is not
clear if this feature is an artifact due to residual structure
from the hot pixels observed in some of the images at this location.
Given the faintness of the
structures, it is difficult to determine if it is due to low-ionisation 
high-velocity material, or if it is an artifact.
Further, our undispersed image is in [O III], so we do not have low-ionisation
undispersed morphology. If the features are real, using the undispersed 
[O III] template, the velocity of recession rises to over 4000 km s$^{-1}$.
Clearly, this would represent a very different kinematic component, and needs
confirmation.

The narrow-band images contain a spot of high signal to the NE and a low signal
spot to the SW, at the position of the `companion'. In our data, no line 
(or continuum) emission is seen from either spot, but the slit width and 
placement probably excluded most of their light.

Line fluxes and ratios were measured (see Table 2). The Balmer decrement 
from H$\alpha$ to H$\beta$ is at least 11, which indicates some reddening.
The ratios of H$\beta$ to [O III] and [N~II] to H$\alpha$ were difficult to measure,
but are close to 0.06 and 0.9, respectively, summing the clouds on either side
of the nucleus. These are
in the range characteristic of ionisation by a Seyfert-type nucleus, rather than
of H II regions or LINER activation. Thus, overall, there is good resemblance
to the kinematic and ionisation behaviour seen in much more detail in the biconical
ejection regions around NGC 4151.

 \section{3C 171}

 This is the highest redshift of the four (0.238) but has large structure
seen optically and in radio. The radio structure is unusual in having
extreme distension of the lobes perpendicular to the inner radio axis.
There is matching [O III] structure (Figure~\ref{fig3}) 
along this inner (10") axis (Heckman 
et al, 1984; Blundell 1996). Heckman et al find [O III] velocities +550 
km s$^{-1}$ to the W and -50 km s$^{-1}$ to the E. The WFPC2 [O III] image 
was rotated and scaled to match the STIS image and used for an undispersed 
template (see Figure~\ref{fig5}). The structure is complex outside the nucleus, 
with the inner optical `jet' not aligned 
with the radio axis. However, the optical structure is bent and does line up 
with the radio on larger scales.

 The STIS G750L spectrum contains both H$\beta$+[O III] and the H$\alpha$ region 
due to the redshift. There is extended line emission in all, and a fairly weak 
nuclear continuum: the continuum cross-section is less peaked than 3C 135, and
asymmetrical (Figure~\ref{fig3}). 
Figure~\ref{fig6} shows extracted spectra from selected regions.
The nuclear spectrum has no broad lines. The emission lines do not peak strongly
at the continuum: the increase is $<$50\%. No off-nuclear continuum is detected. 
We see no sign of any dust lanes.

 The ELR is detected only in the innermost bright knots in the STIS spectrum.
The more distant knots lie outside the slit and were not detected. The inner
knots have clear differential velocities, as in NGC 4151 (and unlike 3C135). 
On the lower (E) side vel is $\sim$-100 km/s and on the upper 
(more extended, W) side, velocities reach 300 km/s. Table 2 and Figure~\ref{vels}
show the velocity measures. Our velocities are in the 
same sense as the Heckman et al (1984) values seen further away from the nucleus.
The brighter side is approaching, although in this case the receding
side is seen in the forbidden lines (only) to greater off-nuclear distances. 
Radio polarisation on the receding (W) side, is consistent with this geometry,
and line of sight. Line ratios in the inner knots are similar to the nucleus
and the nucleus is not a bright point source. Thus, the central engine
may be more hidden in this source than in 3C 135.

\section{3C 78}

 This galaxy is at much lower redshift than the galaxies discussed so far, at
only 8650 km s$^{-1}$. The optical jet is claimed by Sparks et al (1995) to 
be synchrotron, 
based on the exact spatial coincidence of the WFPC2 snapshot image jet with 
the radio jet. These are the only data for that conclusion. 
The narrow-band WFPC2 image obtained later, shows only
weak extended H$\alpha$ emission, which is not associated with the jet, 
and thus consistent with the synchrotron assumption.

The WFPC2 snapshot image has a straight `jet' on one side of the nucleus only.
It is almost normal to the major axis of the central galaxy light, and diverges 
slightly. Light from the jet may be traced to about 2.5 arcsec but is brightest 
in the inner 1
arcsec. The STIS acquisition image shows the jet is lined up well along the slit
direction. Figure~\ref{fig7} shows the profile of the jet at different wavelengths.

 The cross-section across the nucleus is quite asymmetrical in that there is
extra light from the jet seen on both blue and red spectra extending over large
distances. It is seen to 1.5 to 2.0 arcsec in the blue and in the red. The profile
in the red and  blue is similar, but its brightness compared
with the underlying galaxy is higher in the blue. The innermost part is
apparently very blue if we measure it by reflecting the galaxy profile on 
the non-jet side. However (see Fig~\ref{fig7}) this may be caused by dust on the
non-jet side: the non-jet side of the nucleus is fainter in the blue than the red,
suggesting some dust. Spectral plots do not show this to be measurable. 

Spectral plots are shown in Figure~\ref{fig8}. The nuclear emission lines are 
only very slightly extended spatially. The `jet' spectrum has a continuum which is
bluer than the surrounding galaxy, and has some associated faint line emission,
mostly H$\alpha$ and [S II].
The continuum is generally faint but peaks ~6 pixels away from the nucleus, as 
also seen in the undispersed images. There is weak line emission between 
the nucleus and the jet peak flux, which does not follow the 
continuum flux distribution with distance from the nucleus (i.e. is not just
scattered nuclear light). These lines are 
H$\alpha$, [N II] 6584, [S II] 6720, and maybe weak [O~I] 6300 and [O III] 5007.
Cross-correlating these extended emission lines with the nuclear spectrum gives 
shifts that go from -115 km s$^{-1}$ to 0 km s$^{-1}$
at locations from 3 to 6.5 pixels from the nucleus. No line emission is seen 
beyond 7 pixels. No extended line emission is seen in the blue spectrum
([O III] and H$\beta$).

Figure~\ref{fig11} shows profiles of all four galaxies, showing that 3C 78
(and 3C 264) have regions of depressed luminosity around the nucleus, not seen
in the non-jet galaxies, and with radii the same as the jet length. We discuss
this further below.

\section{3C 264}

  This is another nearby galaxy (redshift $\sim$6600 km s$^{-1}$), 
and its optical jet has also been generally claimed as synchrotron, based on 
its spatial concidence with the radio jet (Crane et al 1993).
The jet is very similar in length and flux to that of 3C 78,
and also coincides with the inner radio structure. The overall radio structure
is large, complex, and of type FR I (see e.g. Lara et al 1997). 
It has been referred to as a wide angle tailed source,
and has a weak radio counterjet. There is no previous spectroscopy of the jet.

Baum et al (1997) have published an extensive discussion of the radio and
HST imaging of the jet and central region of the galaxy.
The optical morphology is peculiar in showing the jet to lie within a circular region
of low flux, so that is looks like a pointer on a dial. 
The jet is highly collimated as far as the edge of the ring and then becomes
faint and diffuse. There is weak H$\alpha$ emission within the ring but
not along the jet. 

The luminosity profile shows the inner low flux region clearly (Fig~\ref{fig11}).
The jet profile is also shown in Figure~\ref{fig9}.
The 3C 264 jet has structure that is more marked at shorter wavelengths:
specifically, there are four main knots along its inner length of $\sim$0.7 
arcsec that are brighter at shorter wavelengths. Our spectra give some new 
information on the central region, the nucleus, and the jet.

Figure~\ref{fig10} shows spectral plots.
The jet spectrum has no strong spectral line features: there is weak
H$\alpha$ emission but that is seen in the non-jet side too.
It is difficult to subtract the underlying galaxy spectrum out
of the jet spectrum, but our efforts at this remove the most obvious features:
the 4000A break, the g- and Mg b-band absorptions, and Na D. There seems to be
some H$\alpha$ emission in the galaxy-subtracted jet spectrum.
The galaxy-subtracted jet continuum is bluer than the
galaxy or the nucleus, with the innermost features being the bluest of all.

The low flux region inside the `ring' on the side opposite the jet has the same
colour as the galaxy spectrum outside the ring, and the same spectrum overall.
This indicates that a) the inner region is not dimmed by dust, which would
redden it, as favoured by Baum et al. (1997), or b) the region is completely
obscured by dust so that we see only unreddened stars in front of it.
The latter possibility seems less likely since the jet itself is blue
and would have to be associated with the dust disk or region.
Also, we find from recent WFPC2 colour imaging of ours
that the ring itself is red, and our spectra of
the ring edge (2-3 pixel rows) do show this, compared with spectra from the
surrounding rows. Thus, the phenomenon appears to be that either the
jet itself, or some other nuclear-related process, has cleared out the central
region, and perhaps swept up dust as far as the ring. The collimation
of the inner jet and its disruption at the ring are consistent with this.

The nuclear spectrum has strong emission lines of [O II], [O III], and [N II],
as well as weaker Balmer lines, and [O III] 4363, [S II], and [O I].

\section{Discussion}

  In our investigation of jet-like structures near the nuclei of 3C radio galaxies,
we have found two of each kind of broadly different types of phenomenon: 
`biconical' blobby emission-line region excited by the AGN, and narrow
or well-collimated synchrotron continuum emission from a jet.
In all cases the undispersed broad- and narrow-band imaging offered clues to 
the process. The synchtrotron jets have exact correspondence
with the radio structure and also lie exactly within a region of lowered
luminosity around the galaxy nucleus. In 3C 171, more extended emission 
associated with hot stars (Heckman et al 1984) is very extended
along curved and irregular locations. 

  The line ratios between H$\beta$ and [O III], and H$\alpha$ and [N II] in the
near-nuclear regions of 3C 171 and 3C 135 are poorly defined, but still in the
range ($\sim$0.06 and near 1.0 respectively) that indicate ionisation by a Seyfert-like
source. In the outer regions of 3C 171, where our signal is weaker, the ratios
appear to be similar or more Seyfert-like, corresponding to lower gas density.
These extended line emission regions are seen to be flowing outwards from the 
nucleus at a few hundred km s$^{-1}$, as seen in much more detail in NGC4151. The
outward motions are consistent with the indications from polarisation and flux 
measures in radio structures, and suggest that the line of sight lies outside
the main cones of nuclear illumination, as postulated by unified scenarios for
radio galaxies. In 3C 135 we may see some material moving at several thousand
km s$^{-1}$. This is higher than velocities seen in other objects (up to
1600 km/s in NGC 4151) and so may be associated with the radio-emitting material.

   The 3C 78  and 3C 264 synchrotron jets are blue but otherwise featureless, 
with little or no colour gradients. We see no clear evidence of dust, but 
the jet colours are naturally subject to this possibility.
There is an apparent colour gradient in 3C 273 (see e.g. Hutchings and Neff 1991)
and we will make a detailed comparison later in this program.
The circumnuclear region of low luminosity appears to be cleared of 
material in both cases, as the jets are very collimated and bright until
they reach this boundary. Baum et al (1997) discuss deceleration processes that
are likely at this interface. The region is elliptical in 3C 78 and circular
in 3C 264, which may imply circular regions seen at different inclinations,
or ellipsoidal volumes. If the region has been swept out by the jet, it must
have changed direction rapidly to do so. If it is cleared out by non-jet nuclear 
activity, this may be an essential condition of jet formation. Again, comparison
with the (much larger scale) jet of 3C 273 will be of interest. 
There is some extended line
emission very close to the nucleus that may be a secondary effect of the jet.

  There has been much discussion in recent literature of the interaction
of radio jets with clouds (e.g. Best et al 1998, Lacy et al 1998). In this paper
the two jets visible optically are not associated with clouds, and seem to
interact with the inside wall of a cavity without exciting line emission.
The other two sources contain emission line clouds but there is no clear
indication that the clouds are all excited by radio jets. Thus there appear to be 
different mechanisms at work in different environments. 

  Further observations will be undertaken later in this program. We will
report these and also discuss more detailed models in due course. 

\newpage

\centerline{References}

Baum S. A., O'Dea C.P., Giovannini G., Biretta J., Cotton W.B., De Koff S.,
Feretti L., Golombek D., Lara L., Macchetto F.D., Miley G.K., Sparks W.B.,
Venturi T., Komissarov S.S., 1997, ApJ, 483, 178

Best P.N., Carilli C.L., Garrington S.T., Longair M.S., Rottgering H.J.A., 1998,
MNRAS, in press

Blundell K.M.,1996,  MNRAS, 283, 538

Crane P., et al, 1993, ApJ, 402, L37

De Koff S., Baum S.A., Sparks W.B., Biretta J., Golombek D., Macchetto F.D.,
McCarthy P.J., Miley G.K., 1996, ApJ Supp, 107, 621

Heckman T.M. van Breugel W.J.M., Miley G.K., 1984, ApJ, 286, 509

Hutchings J.B. and Neff S.G., 1991, PASP, 103, 26

Hutchings J.B., et al 1998, ApJ, 492, L115

Lacy M., Rawlings S., Blundell K.M., Ridgway S.E., 1998, MNRAS, in press

Lara L., Cotton W.D., Feretti L., Giovannini G., Venturi T., Marcaide J.M., 
1997, ApJ, 474, 179

Leahy J.P., Black A.R.S., Dennett-Thorpe J., Hardcastle M.J., Komissarov S.,
Perley R.A., Riley J.M., Scheuer P.A.G.,  1997. MNRAS, 291, 20

McCarthy P.J., Spinrad H., van breugel W., 1995, ApJ Supp, 99, 27

Sparks W.B., Golombek D., Baum S.A., Biretta J., De Koff S., Macchetto F.D.,
McCarthy P., Miley G.K., 1995, ApJ, 450, L55

\clearpage

\centerline{\bf Captions to figures}

\figcaption[**.ps]{Images of 3C135. Left: undispersed narrow-band WFPC2 image of 
[O III] line scaled and oriented as the STIS spectral images. 
Centre: dispersed image of H$\beta$ and [O III] lines. Right: dispersed image of
H$\alpha$ + [N II] blend and [S II]. Note the high velocity material in H$\alpha$.
\label{fig1}}

\figcaption[**.ps]{Contour plots of undispersed emission-line images of [O III],
oriented and scaled to match the STIS spectra with the slit edges parallel to the
y-axis. Components measured in different lines in the STIS spectra are labelled. 
The slit width was 0.5 arcsec, and orientation given in Table 1 is vertical.
\label{fig2}}

\figcaption[***.ps]{Cuts along the slit in the spectra of 3C171 and 3C135
at positions of emission lines and the continuum. 3C171 is shown on log scale
as the range of line fluxes is high. The different ionisation of emission line
material can be seen. \label{fig3}}

\figcaption[**.ps]{Spectra from regions in 3C135. The nuclear flux is complete
and the spectrum 
indicates AGN activity. The cut through ABC is shifted by 4e-15 and 
the cut through region FGHI by 5e-15, for clarity. 
The galaxy spectrum is from regions outside the strong
emission lines, but shows weak [O III] and H$\alpha$. 
\label{fig4}}

\figcaption[**.ps]{Measured velocities of emission line features. The labels refer
to measures which identify the features defined in Figure~\ref{fig2}, while the 
unlabelled symbols are from measures
from pixel rows or groups of rows without regard to the labelled features. A' and
I' indicate that these features have more than one cloud, spatially separated,
while two values for B and C indicate spatially overlapping clouds. 
Open and closed symbols
indicate pixel row measures by different authors. The agreement among points and 
symbols indicates the measuring error. 3C 135 presents a more complex situation, and
also has the very high velocities given in Table 2. \label{vels}}

\figcaption[**.ps]{Images of 3C171. Left: undispersed narrow-band WFPC2 image of 
[O III] line scaled and oriented as the STIS spectral images. 
Centre: dispersed image of H$\beta$ and [O III] lines. 
Right: dispersed image of H$\alpha$ + [N II] blend, plus [S II] doublet. 
Note that the outermost line emission seen in the undispersed image lies outside
the slit of the spectral images. Slit orientation from Table 1 is vertical.
\label{fig5}}

\figcaption[**.ps]{Spectra from regions in 3C171. The nuclear spectrum is shifted
by 6.e-15, the cuts through region D by 4.5e-15, and the cut
through region B by 2.5e-15. The galaxy spectrum is from regions outside the
emission lines. Note the weak H emission in C/D and the lack of Mg-b absorption,
indicating lack of old stellar population. \label{fig6}}

\figcaption[**.ps]{Cuts along the slit in the spectra of 3C78 in line-free regions
from G450L and G750L. Top and centre: The jet lies along
the slit to the left of the nucleus in the solid plots. The dotted profiles are
reflected about the nucleus and the difference (dashed line) shows the approximate 
flux along the jet. Bottom: the cuts normalised to the outer galaxy signal from
both gratings. The inner jet is bluer while there is a redder region (dust?) on
the non-jet side. Thus, the reflected profile in the upper panels may be affected
by this. \label{fig7}}

\figcaption[**.ps]{Spectra from regions in 3C78, from both gratings. The total nuclear 
spectrum flux is shown, but the other extractions are from adding rows in
the spectral images. Note that the jet is bluer than the galaxy (region just outside
the jet radius on the opposite side). The `jet-galaxy' spectrum subtraction was
done to remove the stellar features from the jet spectrum. The result is blue 
and featureless except for H$\alpha$ and [S II] emission, but has a red tail. 
The `hole' spectrum is from the region opposite 
the jet but at the same radii, and is similar to the jet. Note that the H$\alpha$
and [S II] lines are extended around the nucleus. \label{fig8}}

\figcaption[**.ps]{Azimuthally averaged profiles of the broad-band images of 
all galaxies with spatial scales corresponding to the redshift of 3C78. 
The jets have been roughly edited out of 3C78 and 3C264 and the fluxes along 
the jets alone dotted in. `Jet' fluxes contribute little to the other two galaxies. 
Note that both the jet galaxies show a region of lower flux around the nucleus,
ending where the jet flux drops. This is not seen in the non-jet galaxies.
\label{fig11}}

 \figcaption[**.ps]{Cuts along the slit in the spectra of 3C264 in line-free regions
from G450L and G750L. Top and centre: The jet lies along
the slit to the left of the nucleus in the solid plots. The dotted profiles are
reflected about the nucleus and the difference (dashed line) shows the approximate 
flux along the jet. The jet becomes increasingly red with distance from the
nucleus. Bottom: the cuts normalised to the outer galaxy signal from
both gratings. The region opposite the jet (inside the ring) is almost
the same colour as the outer galaxy, indicating that the inner ring is not dusty.
\label{fig9}}

\figcaption[**.ps]{Spectra from regions in 3C264, from both gratings. The nuclear 
spectrum flux is scaled by 1.5 for clarity, and other extractions are from adding 
selected rows in the spectral images. The galaxy  spectrum is from outside
the ring opposite the jet and the hole is inside the ring opposite the jet. 
The `jet-galaxy' spectrum subtraction was
done to remove the stellar features from the jet spectrum. The result is blue 
and featureless, except for weak H$\alpha$ emission. 
The `hole' spectrum is identical to the galaxy spectrum and is
displaced by 3.E-15 for clarity. Note that the H$\alpha$
emission is extended around the nucleus. \label{fig10}}

\clearpage

\begin{deluxetable}{lclccc}
\tablenum{1}
\tablecaption{STIS observations of 3C galaxies}
\tablehead{\colhead{Name} & \colhead{z} &\colhead{1997 Date} 
&\colhead{Slit angle\tablenotemark{a}} &\colhead{Grating} &\colhead{Exp (sec)}}
\startdata
3C 78  & 0.029 & Oct 14 1997 & 55$^o$ &G430L & 3070\nl
&&&&G750L & 1240 \nl
3C 135 & 0.127 & Oct 7 1997 & 51$^o$ &G750L & 2020\nl
3C 171 & 0.238 & Sep 20 1997 & 60$^o$ &G750L & 2300\nl
3C 264 & 0.022 & Jan 6 1998 & 44$^o$ &G430l & 1180\nl
&&&&G750L &790\nl
\enddata
\tablenotetext{a}{Slit orientation in degrees E of N-S: dispersion normal to this}
\end{deluxetable}

\clearpage

\begin{deluxetable}{llcl}
\tablenum{2}
\tablecaption{Velocities and fluxes of emission-line clouds}
\tablehead{\colhead{Feature} &\colhead{Velocity} &\colhead{[O III]} 
&\colhead{H$\alpha$}\\
&\colhead{km s$^{-1}$} &\multicolumn{2}{c}{10$^{-14}$ erg cm$^{-2}$ sec$^{-1}$}}
\startdata
{\bf ~~~~~~~3C 135}\nl
A & 300 to 530 & 3.7 & 1.8 \nl
B & 100, 300 & 2.5 & 2.5(B+C) \nl
C & -100, 300 &  1.8 & -- \nl
D & 20 &  6.0 & 4.1 \nl
X & 0 &  2.6 & -- \nl
E & 2500 to 4300? & -- & 0.2 \nl
F+G & 160 & 5.7 & 5.7 \nl
H+I & 420, 0, -200 & 3.9 & 2.8 \nl
\hline
{\bf ~~~~~~~3C 171}\nl
B & -80 to -150 & 6.2 &  3.2 \nl
X & 0  & 4.5 & 6.0 \nl
C & 100 to 300 & 4.8 & 2.4 \nl
D & 90 to 160 & 1.0 & 0.7 \nl
\enddata
\end{deluxetable}

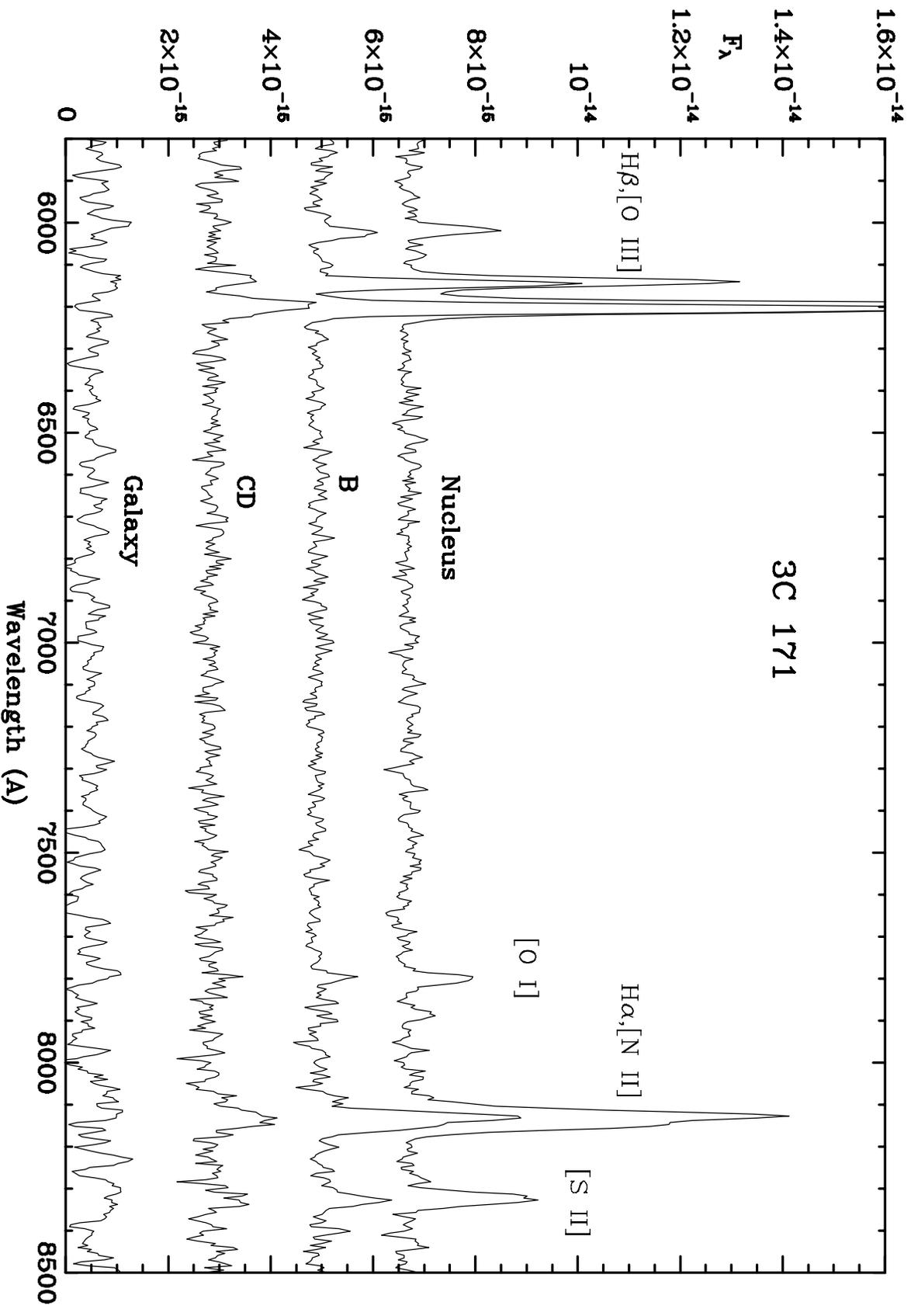
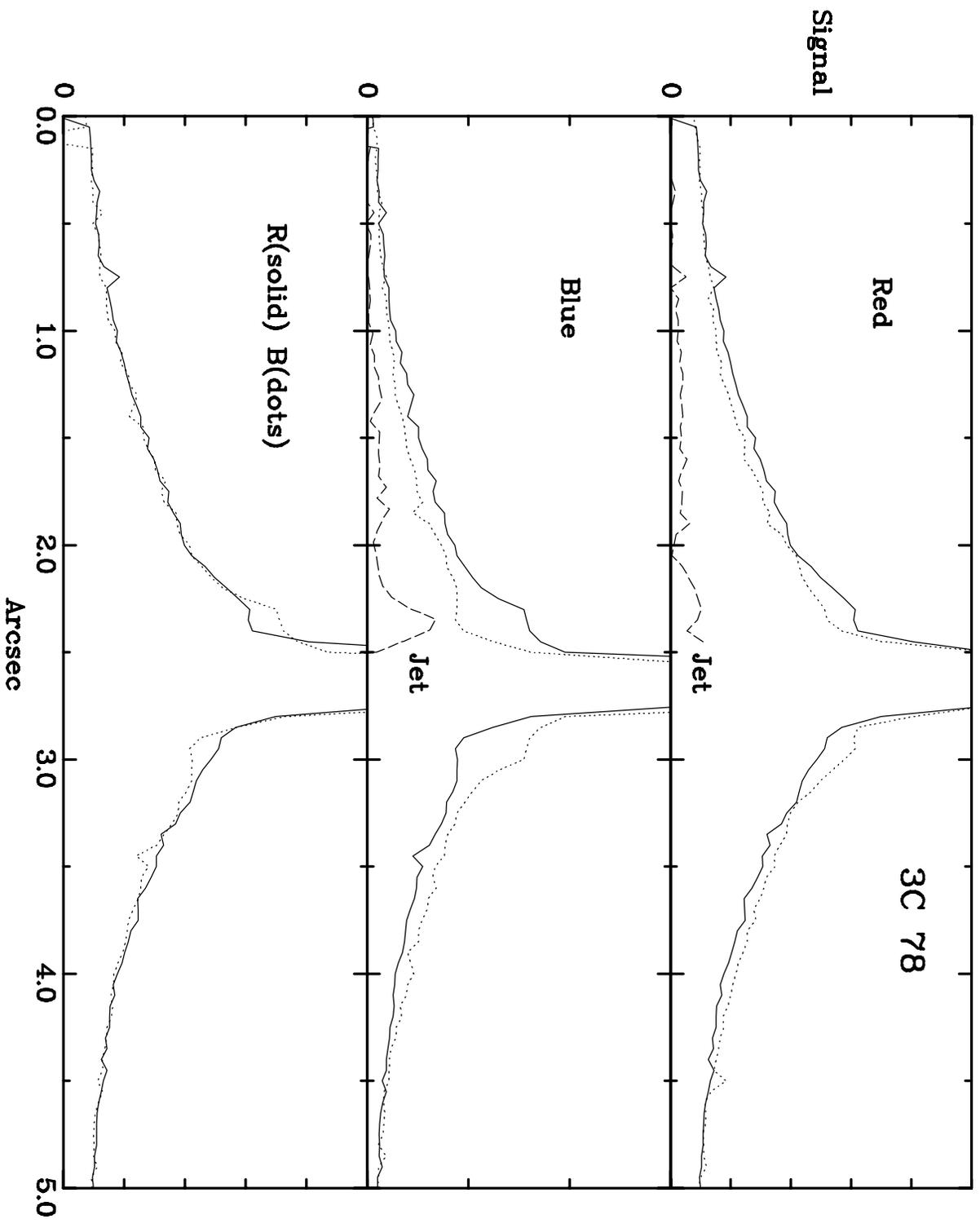
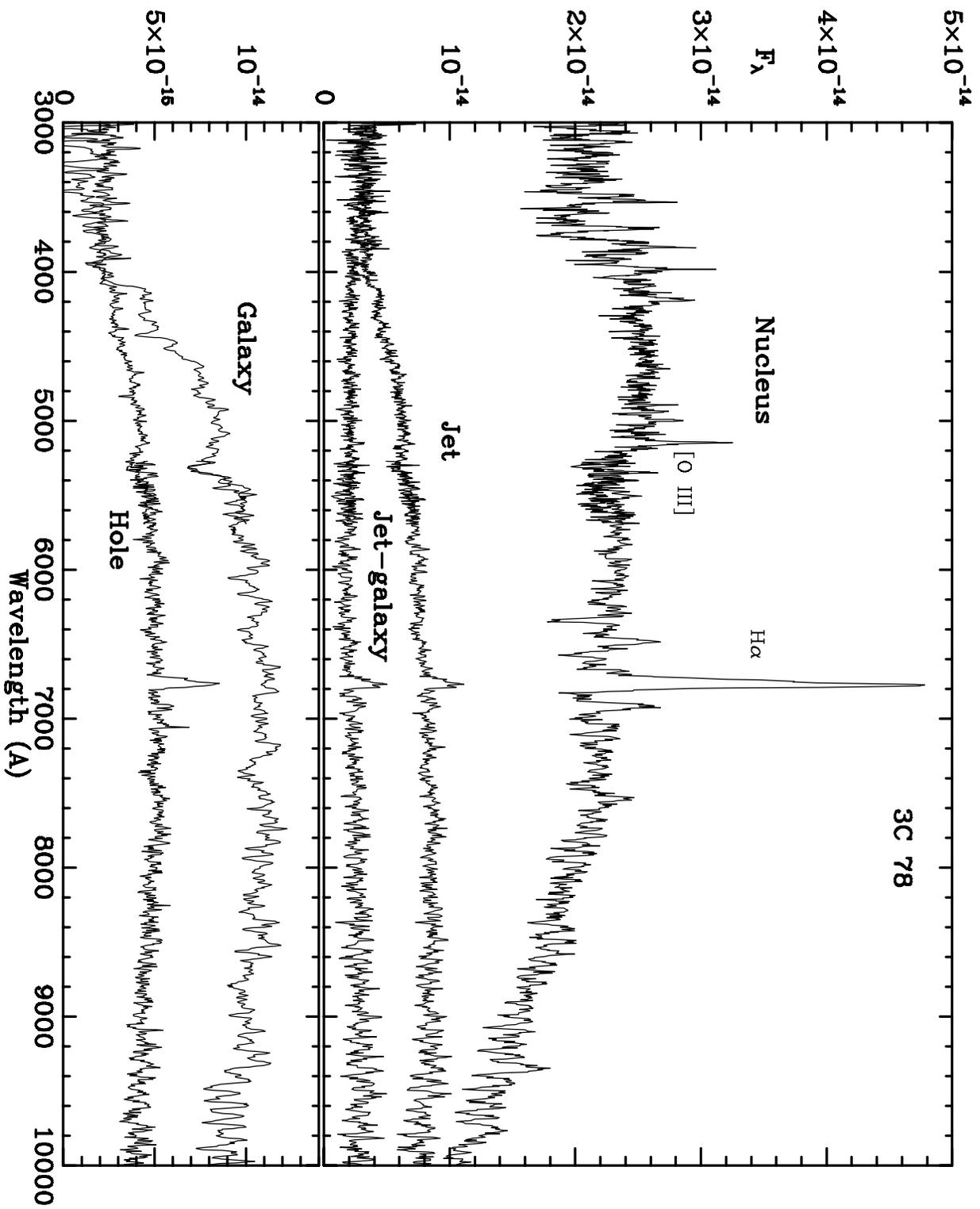
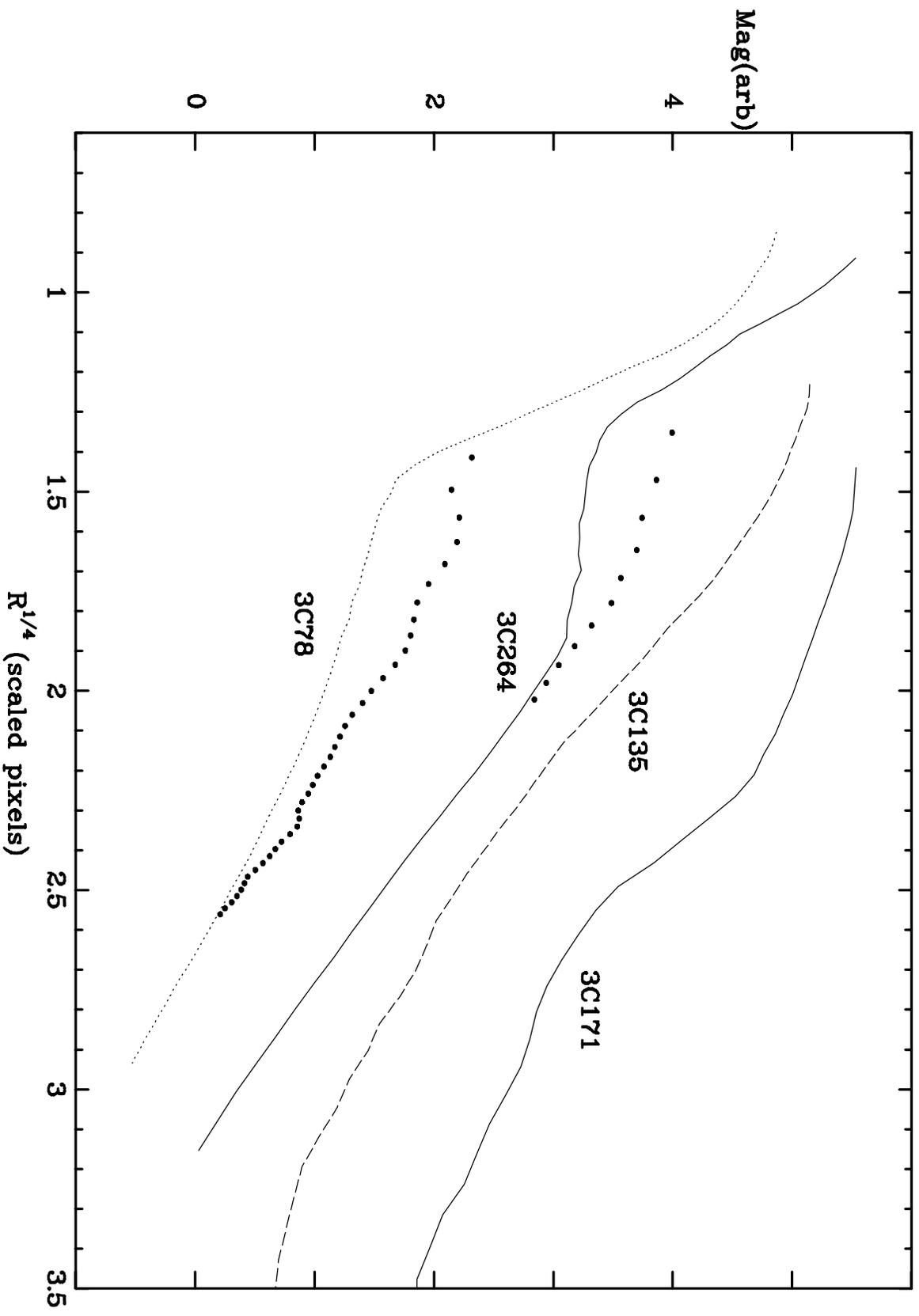
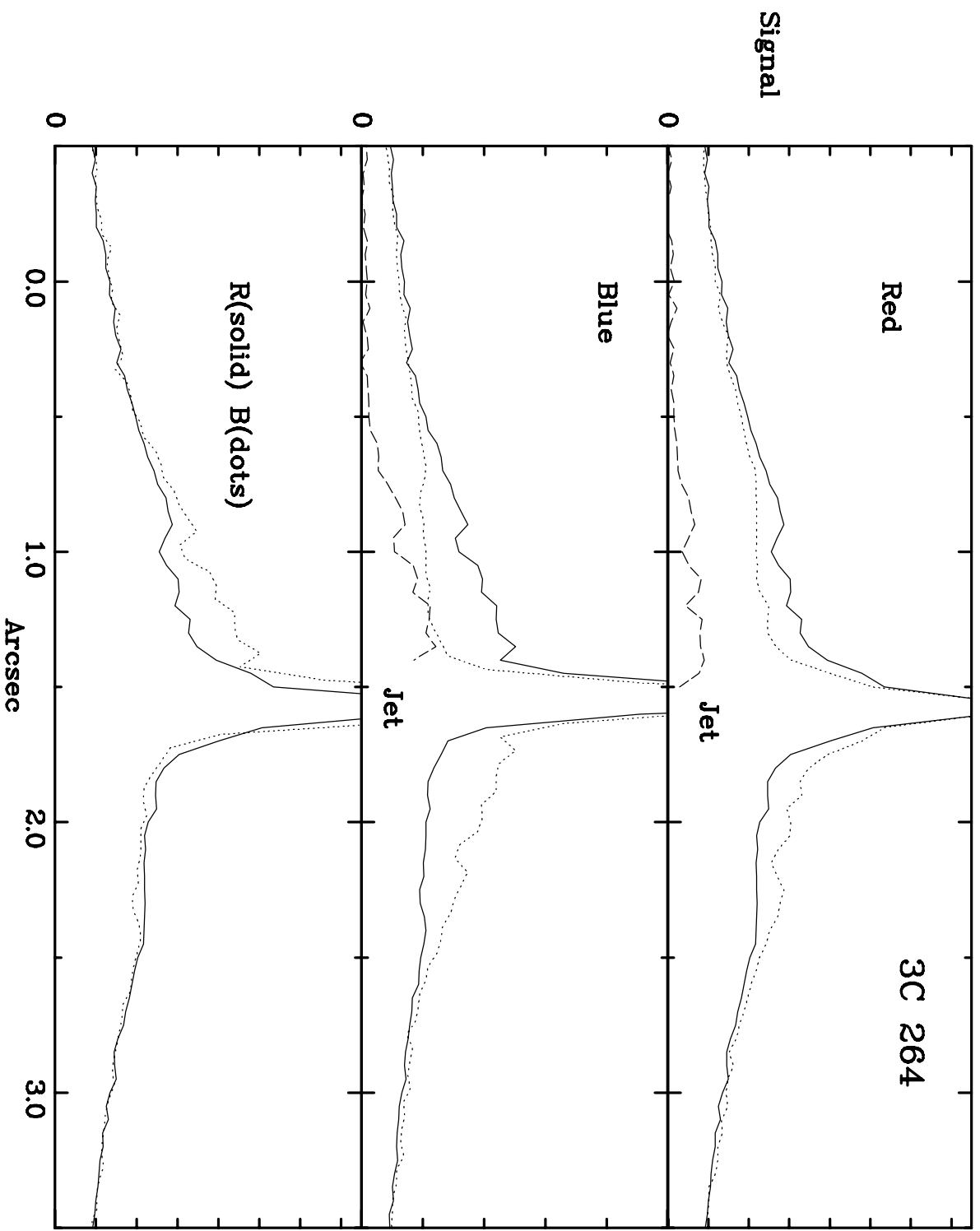
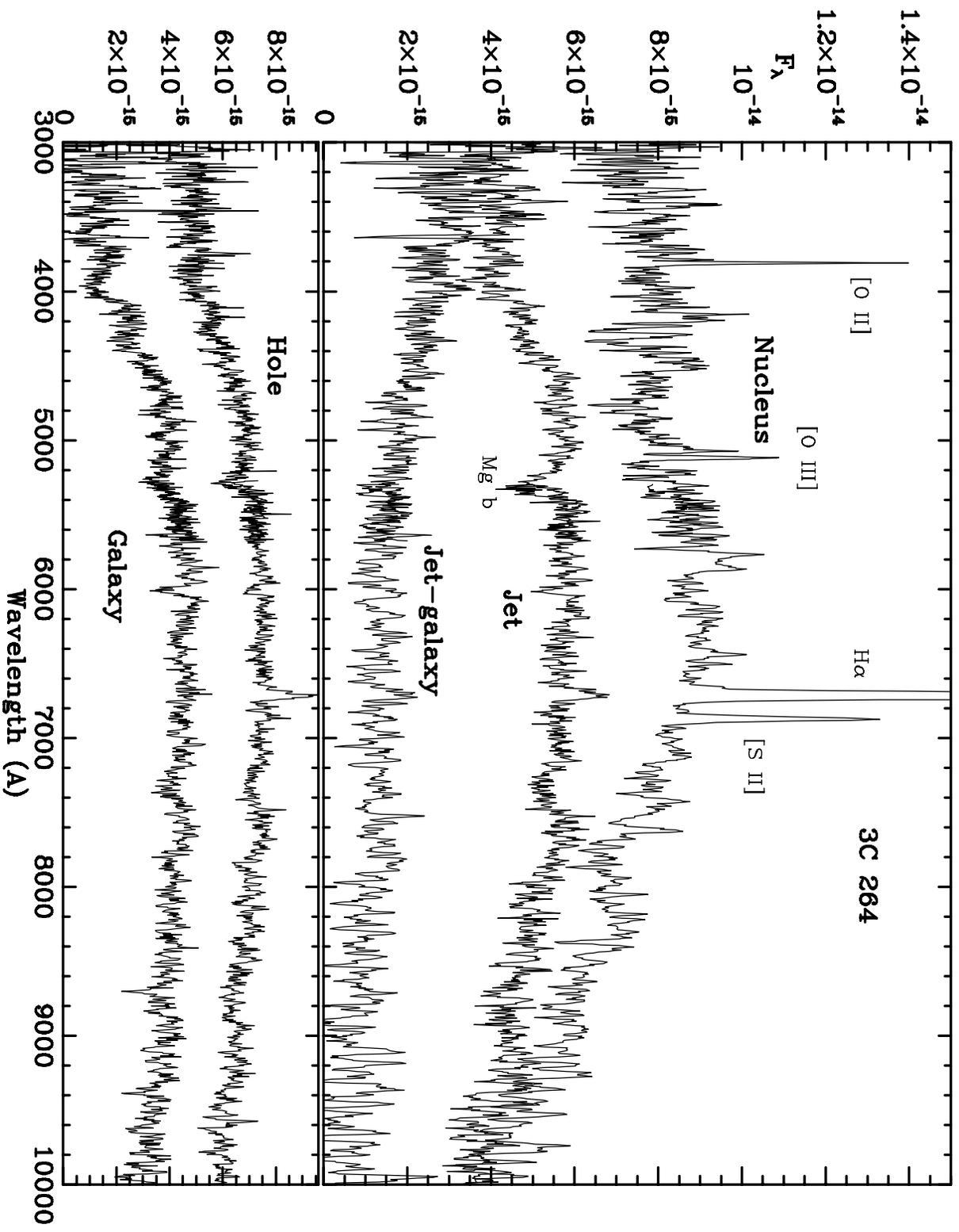
\end{document}